# Exploring a cold plasma - 2d black hole connection

Floyd L. Williams

Department of Mathematics and Statistics
University of Massachusetts, Amherst, MA 01003
Email: williams@math.umass.edu

## Abstract

Using a resonance nonlinear Schrödinger equation as a bridge, we explore a direct connection of cold plasma physics to two-dimensional black holes. Namely, we compute and diagonalize a metric attached to the propagation of magneto-acoustic waves in a cold plasma subject to a transverse magnetic field, and we construct an explicit change of variables by which this metric is transformed exactly to a Jackiw-Teitelboim black hole metric.

## 1  Introduction

In the closing remarks in [1] we indicated briefly a connection of black holes in the Jackiw-Teitelboim model of two-dimensional dilaton gravity to the dynamics of two-component cold collisionless plasma in the presence of an external transverse magnetic field. The purpose of the present paper is to greatly expand those brief remarks in various directions. For example, we explore this connection for plasma metrics derived more generally from Gurevich-Krylov solutions of the associated magnetoacoustic system (MAS) [2]

$$\rho_{1\,t} + (\rho_1 u_1)_x = 0$$
$$u_{1\,t} + u_1 u_{1\,x} + \rho_{1\,x} + (\delta - 1)\left[\frac{\rho_{1\,xx}}{\rho_1} - \frac{1}{2}\left(\frac{\rho_{1x}}{\rho_1}\right)^2\right]_x = 0, \tag{1}$$

for $\delta > 1$, which describes the uni-axial propagation of long magnetoacoustic waves in a cold plasma of density $\rho_1(x,t) > 0$ with velocity $u_1(x,t)$ across a magnetic field. Moreover, we present a concrete description of this connection.

We start with a *resonance* nonlinear Schrödinger (RNLS) equation [3], [4], [5]

$$i\Psi_t + \Psi_{xx} + \gamma|\Psi|^2\Psi = \delta\frac{|\Psi|_{xx}}{|\Psi|}\Psi. \tag{2}$$

Here $\frac{|\Psi|_{xx}}{|\Psi|}$ is a de Broglie quantum potential, and $\gamma, \delta$ are real numbers with $\delta > 1$. This $\delta$ we will take to be the same as that in the system (1); $i^2 = -1$.

Solutions of the form $\Psi = e^{R-iS}$ for real-valued functions $R(x,t), S(x,t)$ are considered. Since $R, S$ are real, it follows directly that equation (2) is equivalent to the system of equations (= Madelung fluid equations)

$$R_t - S_{xx} - 2R_x S_x = 0$$
$$S_t + (1-\delta)R_{xx} + (1-\delta)R_x^2 - S_x^2 + \gamma e^{2R} = 0. \tag{3}$$

In Section 2, we review (or slightly generalize, for the sake of completeness) a result in [4], [5] that such solutions $R, S$ (or therefore $\Psi$ in (2)) are in correspondence with solutions $\rho_1 > 0$, $u_1$ of the system (1), in case $\gamma < 0$. An approach to the latter system by way of a shallow wave approximation also appears in [4], [5]. In those references $\gamma = -\frac{1}{2}$, and the parameter $B$ there

corresponds to our $\sqrt{\delta-1}$. From (3) it follows that for

$$r(x,t) \stackrel{def}{=} exp(R(x,\tfrac{t}{\sqrt{\delta-1}}) + \tfrac{S(x,\tfrac{t}{\sqrt{\delta-1}})}{\sqrt{\delta-1}})$$
$$s(x,t) \stackrel{def}{=} -exp(R(x,\tfrac{t}{\sqrt{\delta-1}}) - \tfrac{S(x,\tfrac{t}{\sqrt{\delta-1}})}{\sqrt{\delta-1}}) \qquad (4)$$

the reaction diffusion system (RDS)

$$r_t - r_{xx} + Br^2 s = 0$$
$$s_t + s_{xx} - Brs^2 = 0 \qquad (5)$$

is solved for the value $B = \tfrac{-\gamma}{(\delta-1)}$. $r,s$ here correspond to $e^{(+)}, e^{(-)}$ respectively in the preceding references where again $\gamma = -\tfrac{1}{2}$ there. Also see [6]. For $\delta > 1$, the RNLS equation is not reducible to a nonlinear Schrödinger equation but instead to a RDS.

Now given the RDS in (5), the point for us is that one can construct from its solutions $r,s$ a pseudo Riemannian metric

$$g: ds^2 = g_{11}dt^2 + 2g_{12}\, dt\, dx + g_{22}\, dx^2,$$
$$g_{11} \stackrel{def}{=} -r_x s_x, \quad g_{12} \stackrel{def}{=} \tfrac{1}{2}(sr_x - rs_x), \quad g_{22} \stackrel{def}{=} rs \qquad (6)$$

of *constant* Ricci scalar curvature $R(g) = 4B$ $(= \tfrac{-4\gamma}{(\delta-1)}$ in our case); see [3], [7], [8]. Constant curvature is a required ingredient for the J-T (Jackiw-Teitelboim) theory of 2d gravity [9],[10],[11],[12]. Since $r,s$ are defined in terms of the solutions $R,S$, which are in correspondence with solutions $\rho_1, u_1$ of (1) (as we have noted) $g$ also will correspond to solutions $\rho_1, u_1$ of the MAS (1). Thus we will also denote $g$ by $g_{plasma}$. Details of solutions $\rho_1, u_1$ (as traveling waves expressed in terms of the Jacobi elliptic function $dn(x,\kappa)$) and a computation of the metric $g = g_{plasma}$ in terms of them are provided in Section 2. In general of course $g$ is non-diagonal: $g_{12}(= g_{21}) \neq 0$ in (6). In Section 3 we establish integrability conditions involving $\delta$ and parameters defined in the solutions $\rho_1, u_1$ that suffice for the existence of a change of variables by which $g$ assumes a simpler diagonal form.

The main results are presented in Section 4. There we provide another explicit change of variables that transforms $g_{plasma}$ precisely into a (Lorentzian) J-T black hole metric $g_{bh}$ – which therefore explicates the proposed cold plasma – black hole connection. Using this same transformation, we construct an explicit dilaton $\Phi_{plasma}$ such that the pair $(g_{plasma}, \Phi_{plasma})$ solves the J-T gravitational field equations – equations that involve a cosmological constant $\Lambda$, which is shown to have the value $\tfrac{2\gamma}{(\delta-1)}$ for the $\gamma$ and $\delta$ in (2), this $\delta$ being the same $\delta$ in (1). Another plasma-black hole connection revealed in Section 4 is the observation that the Hawking black hole temperature and entropy can be expressed in terms of parameters involved in the description of solutions $\rho_1, u_1$ of the plasma system (1).

## 2   Formulas for the cold plasma metric

As mentioned in Section 1, a correspondence between solutions of the system (1) and the system

(3) (or, equivalently, of the RNLS equation (2)) will be reviewed in this section, under the assumption that $\gamma < 0$. We also find an initial, general formula for the plasma metric; see (11). A concrete formula then follows as concrete solutions of the system (1) are considered. The end result is given in (17).

For the correspondence, one direction is quite straight forward: Given solutions $R, S$ of (3), define
$$\rho_1 = -2\gamma e^{2R} > 0, \quad u_1 = -2S_x \tag{7}$$
for $\gamma < 0$. Then one can check that the equations in (1) follow. Conversely, suppose $\rho_1 > 0, u_1$ are solutions of the system (1). One should clearly define
$$R = \frac{1}{2} \log\left(\frac{\rho_1}{-2\gamma}\right) \tag{8}$$
so that $\rho_1 = -2\gamma e^{2R} > 0$ as in (7). For the next step, first choose any $S_0$ such that $-2S_{0_x} = u_1$. Then the first equation in (1) leads to the first equation in (3) for the pair $R, S_0$, and the second equation in (1) implies that the partial derivative of
$$-S_{0_t} + S_{0_x}^2 - \gamma e^{2R} + (\delta - 1)(R_{xx} + R_x^2)$$
with respect to $x$ vanishes. Thus this quantity is a function $\phi(t)$ of $t$ only. Choose $h(t)$ such that $h'(t) = \phi(t)$ and define
$$S(x, t) = S_0(x, t) + h(t). \tag{9}$$
Then the pair $R, S$ in (8), (9) satisfies both equations in (3), and the proposed correspondence $(\rho_1 > 0, u_1) \leftrightarrow (R, S)$ is established for $\gamma < 0$.

Given the definition of $r, s$ in (4), and the prescription for $g$ in (6), we arrive at the following formulas, where we set
$$\beta \stackrel{def}{=} +\sqrt{\delta - 1}:$$
$$\begin{aligned} g_{22}(x, t) &= -e^{2R(x, t/\beta)}, \\ g_{12}(x, t) &= \frac{-1}{\beta} e^{2R(x, t/\beta)} S_x(x, t/\beta) \\ g_{11}(x, t) &= e^{2R(x, t/\beta)} [R_x^2 - \frac{1}{\beta^2} S_x^2](x, t/\beta) \\ &= e^{2R(x, t/\beta)} [\frac{S_t}{-\beta^2} + R_{xx} + 2R_x^2 + \frac{\gamma}{-\beta^2} e^{2R}](x, t/\beta). \end{aligned} \tag{10}$$
The second expression for $g_{11}(x, t)$ here follows from the second equation in (3). By way of the correspondence $(\rho_1 > 0, u_1) \leftrightarrow (R, S)$ just established, we can also write (for $\gamma < 0$, which we now assume throughout)
$$\begin{aligned} g_{22}(x, t) &= \frac{\rho_1(x, t/\beta)}{2\gamma} \\ g_{12}(x, t) &= \frac{\rho_1(x, t/\beta) u_1(x, t/\beta)}{(-4\gamma\beta)} \\ g_{11}(x, t) &= \frac{\rho_1(x, t/\beta)}{-2\gamma} [\frac{1}{4}(\frac{\rho_{1,x}}{\rho_1})^2(x, t/\beta) - \frac{1}{4\beta^2} u_1^2(x, t/\beta)]. \end{aligned} \tag{11}$$
$R$ is given uniquely in terms of $\rho_1$ by definition (8). However, there could be many other choices for $S$ in (9) for which the pair $(R, S)$ solves the system (3), and for which $u_1 = -2S_x$ (as in (7)). We assert that the metric $g$ is *not* affected by another such choice $S'$, in place of $S$. Indeed,
$$-2S'_x = u_1 = -2S_x \Rightarrow S'_x = S_x,$$
so the assertion follows by (10). Note also that
$$(S' - S)_x = 0 \Rightarrow S'(x, t) = S(x, t) + c(t)$$

for some function of integration $c(t)$. $S'_{xx} = S_{xx}$ (since already $S'_x = S_x$), and $S'_t = S_t + c'(t)$. But by the second equation in (3) for $(R, S')$, and then also for $(R, S)$
$$0 = S_t + c'(t) + (1-\delta)[R_{xx} + R_x^2] - S_x^2 + \gamma e^{2R} = c'(t) \Rightarrow c(t) = c,$$
for some constant 'c'. Thus in fact we see also that $S'_t = S_t$, and $S' = S + c$.

The goal now is to express the plasma metric $g$ more concretely in terms of concrete solutions $\rho_1 > 0, u_1$ of the MAS (1). From [2], traveling wave solutions are given by choosing $\alpha_3 > \alpha_2 \geq \alpha_1 \geq 0, u_0 > 0$, and setting

$$\rho_1(x,t) \stackrel{def}{=} \alpha_1 + (\alpha_3 - \alpha_1) \, dn^2(\frac{(\alpha_3-\alpha_1)^{1/2}}{2} \frac{(x-u_0 t)}{\beta}, \kappa)$$

$$u_1(x,t) \stackrel{def}{=} u_0 + \frac{C}{\rho_1(x,t)}; \qquad (12)$$

$$C = +(\alpha_1 \alpha_2 \alpha_3)^{1/2}, \qquad \kappa = (\frac{\alpha_3-\alpha_2}{\alpha_3-\alpha_1})^{1/2}.$$

We could also replace $u_0 + \frac{C}{\rho_1}$ here by $u_0 - \frac{C}{\rho_1}$. $dn(x, \kappa)$ denotes a standard Jacobi elliptic function with elliptic modulus $\kappa$ [13], and $\beta \stackrel{def}{=} +\sqrt{\delta - 1} > 0$ by (10). For our purpose it is convenient to set

$$a_0 \stackrel{def}{=} +\frac{(\alpha_3-\alpha_1)^{1/2}}{2\beta} > 0, \qquad v \stackrel{def}{=} \frac{u_0}{\beta} > 0 \qquad (13)$$

and therefore write

$$\rho_1(x,t) = \alpha_1 + 4a_0^2 \beta^2 \, dn^2(a_0(x - \beta v t), \kappa). \qquad (14)$$

It is also important to note that we can write

$$C = \alpha_1^{1/2} [4a_0^2 \beta^2 (1 - \kappa^2) + \alpha_1]^{1/2} [4a_0^2 \beta^2 + \alpha_1]^{1/2}. \qquad (15)$$

Namely,

$$1 - \kappa^2 \stackrel{def}{=} 1 - \frac{(\alpha_3-\alpha_2)}{(\alpha_3-\alpha_1)} = \frac{(\alpha_2-\alpha_1)}{(\alpha_3-\alpha_1)}$$

and (again)

$$4a_0^2 \beta^2 \stackrel{def}{=} \alpha_3 - \alpha_1 \quad 4a_0^2 \beta^2 (1-\kappa^2) + \alpha_1 = \alpha_2 \text{ and } 4a_0^2 \beta^2 + \alpha_1 = \alpha_3$$

implies the right hand side in (15) is $\alpha_1^{1/2} \alpha_2^{1/2} \alpha_3^{1/2} \stackrel{def}{=} C$.

Some basic facts regarding the standard Jacobi elliptic functions $sn(x, \kappa), cn(x, \kappa), dn(x, \kappa)$ (for any elliptic modulus $\kappa$) are summed up as follows [13]:

$$sn^2(x, \kappa) + cn^2(x, \kappa) = 1, \qquad dn^2(x, \kappa) + \kappa^2 \, sn^2(x, \kappa) = 1,$$
$$\frac{d}{dx} sn(x, \kappa) = cn(x, \kappa) \, dn(x, \kappa), \qquad \frac{d}{dx} cn(x, \kappa) = -sn(x, \kappa) \, dn(x, \kappa), \qquad (16)$$
$$\frac{d}{dx} dn(x, \kappa) = -\kappa^2 \, sn(x, \kappa) \, cn(x, \kappa),$$
$$sn(x, 1) = tanh(x), \qquad cn(x, 1) = dn(x, 1) = sech(x).$$

Applying the formulas in (11) to $u_1, \rho_1$ given in (12), (14) ($g_{11}$ obviously being the main thing to compute) one arrives at the following concrete formulas for the plasma metric:

$$g_{22}(x,t) = \frac{[\alpha_1 + 4a_0^2 \beta^2 \, dn^2(a_0(x-vt),\kappa)]}{2\gamma}$$

$$g_{12}(x,t) = \frac{-v}{\gamma} a_0^2 \beta^2 \, dn^2(a_0(x-vt),\kappa) - \frac{v\alpha_1}{4\gamma} - \frac{C}{4\gamma\beta}$$

$$g_{11}(x,t) = 4a_0^2 \beta^2 \left[-\frac{a_0^2 \kappa^4}{2\gamma}(sn^2 \, cn^2)(a_0(x-vt),\kappa) + \frac{v^2}{8\gamma} dn^2(a_0(x-vt),\kappa)\right] \quad (17)$$

$$+ \frac{16\alpha_1 \kappa^4 a_0^4 \beta^2 (sn^2 \, cn^2)(a_0(x-vt),\kappa) + \frac{C^2}{\beta^2}}{8\gamma[\alpha_1 + 4a_0^2 \beta^2 \, dn^2(a_0(x-vt),\kappa)]} + \frac{v^2 \alpha_1}{8\gamma} + \frac{vC}{4\gamma\beta},$$

where we use (16) to get that $\rho_{1x}^2(x,t) = 64\kappa^4 a_0^6 \beta^4 (sn \, cn \, dn)^2 (a_0(x-\beta vt),\kappa)$, or

$$\rho_{1x}^2(x,t/\beta) \overset{def}{=} 64\kappa^4 a_0^6 \beta^4 (sn \, cn \, dn)^2 (a_0(x-vt),\kappa), \quad (18)$$

and where $v \overset{def}{=} u_0/\beta$ (by (13)), and $C$ is given by the formula (15). From Section 1 we know that this metric has constant Ricci scalar curvature

$$R(g) = \frac{-4\gamma}{\beta^2} \overset{def}{=} \frac{-4\gamma}{(\delta-1)}.$$

Our convention for scalar curvature is spelled out on page 182 of [12], for example. For other authors, see [9] for example, there is a *sign* difference: Our $R(g)$ would correspond to their $-R(g)$.

For plasma physics a relevant choice for $\alpha_1, \alpha_2$ is the value 1, so that the plasma density $\rho_1$ achieves the convenient value 1 as $|x| \to \infty$: $\kappa = 1$ by (12) implies

$$\rho_1(x,t) = 1 + 4a_0^2 \beta^2 \, sech^2(a_0(x-\beta vt)) \quad (19)$$

by (14), (16). Similarly, the elliptic functions $sn, cn$ in the formulas (17) simplify as hyperbolic functions for $\kappa = 1$. For $\alpha_1 = \kappa = 1, C = (4a_0^2\beta^2 + 1)^{1/2}$, for example, by (15).

## 3   Diagonalization of the plasma metric

In this section we focus on the existence of a change of variables that diagonalizes $g = g_{plasma}$ in (17). Such a simplified version of $g$ would be of quite an advantage as a goal is to eventually map $g$ to a black hole metric. It will be shown that the two conditions spelled out in (33) below (that simply require $v^2$ to be sufficiently large) suffice to insure such a diagonalization. These conditions are prototypical in the sense that similar ones will be set up in Section 4 to insure that $g$ indeed is mapped to a black hole metric.

Consider the initial change of variables $(x,t) \to (\rho,t)$ for $\rho \overset{def}{=} a_0(x-vt)$.

$$x = \frac{\rho}{a_0} + vt \Rightarrow dx = \frac{d\rho}{a_0} + vdt, \quad dx^2 = \frac{d\rho^2}{a_0^2} + \frac{2v}{a_0} d\rho dt + v^2 dt^2,$$

$$dxdt = \frac{d\rho dt}{a_0} + vdt^2.$$

Then by (6),

$$g = (g_{11} + 2vg_{12} + v^2 g_{22})dt^2 + \frac{2}{a_0}(g_{12} + vg_{22})d\rho dt + \frac{g_{22}}{a_0^2}d\rho^2,$$

the point being that

$$g_{ij}(x,t) = g_{ij}(\frac{\rho}{a_0} + vt, t)$$

here depends only on $\rho$ (not on $t$) by the formulas in (17): We can therefore write

$$g = A(\rho)dt^2 + C_1(\rho)d\rho\, dt + C_2(\rho)d\rho^2:$$
$$A(\rho) \stackrel{def}{=} (g_{11} + 2vg_{12} + v^2 g_{22})(\rho),$$
$$C_1(\rho) \stackrel{def}{=} \frac{2}{a_0}(g_{12} + vg_{22})(\rho), \tag{20}$$
$$C_2(\rho) \stackrel{def}{=} \frac{g_{22}(\rho)}{a_0^2}.$$

It follows by Section 2 of [1] (or by a direct, independent argument) that *if* there exists a function $\phi(\rho)$ such that
$$\phi'(\rho) = \frac{C_1(\rho)}{2A(\rho)}$$
(an integrability condition, as was referenced in the introduction), then the change of variables $\tau = t + \phi(\rho)$ reduces $g$ to the diagonal form
$$g = A(\rho)d\tau^2 + [C_2(\rho) - \frac{C_1^2(\rho)}{4A(\rho)}]d\rho^2. \tag{21}$$

By (20),
$$\frac{C_1^2(\rho)}{4} - A(\rho)C_2(\rho) = \frac{(g_{12}^2 - g_{11}g_{22})(\rho)}{a_0^2} = \frac{-(\det g)(\rho)}{a_0^2}$$

and since
$$C_2 - \frac{C_1^2}{4A} = -\frac{[\frac{C_1^2}{4} - AC_2]}{A},$$

equation (21) can be written as
$$g = A(\rho)d\tau^2 - \frac{[\frac{-(\det g)(\rho)}{a_0^2}]}{A(\rho)}d\rho^2. \tag{22}$$

Now $C_1(\rho)$, $A(\rho)$ in particular are continuous functions (they are actually $C^\infty$ functions) so that if $A(\rho)$ is non-vanishing the integrability condition $\phi'(\rho) = \frac{C_1(\rho)}{2A(\rho)}$ can be satisfied. We have assumed that $A(\rho) \neq 0$ of course in deriving (22) — an assumption that we shall explore presently.

From (17), $(2vg_{12} + v^2 g_{22})(\rho)$ reduces to the single term $\frac{-2vC}{4\gamma\beta}$. Therefore by (20) and (17) again
$$A(\rho) = g_{11}(\rho) - \frac{2vC}{4\gamma\beta} = 4a_0^2\beta^2[\frac{-a_0^2\kappa^4}{2\gamma}sn^2(\rho,\kappa)cn^2(\rho,\kappa) + \frac{v^2}{8\gamma}dn^2(\rho,\kappa)] \tag{23}$$
$$+ \frac{16\alpha_1\beta^2\kappa^4 a_0^4 sn^2(\rho,\kappa)cn^2(\rho,\kappa) + \frac{C^2}{\beta^2}}{8\gamma[\alpha_1 + 4a_0^2\beta^2 dn^2(\rho,\kappa)]} + \frac{v^2\alpha_1}{8\gamma} - \frac{vC}{4\gamma\beta}.$$

One could use (17) a third time, or more simply use (11) to compute
$$(g_{11}g_{22} - g_{12}^2)(\rho) = \frac{-\rho_{1x}^2(x,t/\beta)}{16\gamma^2}$$
and hence conclude by (18) that
$$(\det g)(\rho) = \frac{-4\kappa^4 a_0^6 \beta^4}{\gamma^2}(sn^2 cn^2 dn^2)(\rho,\kappa). \tag{24}$$

Equation (22) then assumes the concrete form
$$g = A(\rho)d\tau^2 - [\frac{1}{A(\rho)}\frac{4\kappa^4 a_0^4 \beta^4}{\gamma^2}(sn^2 cn^2 dn^2)(\rho,\kappa)]d\rho^2 \tag{25}$$
for $A(\rho)$ given by (23). In the special case when $\alpha_1$ is chosen to be 0, we get $C = 0$

(by definition (12)) so that $A(\rho)$ simplifies greatly to
$$A(\rho) = 4a_0^2\beta^2 dn^2(\rho,\kappa)[a_0^2\kappa^4(\tfrac{sn^2 cn^2}{dn^2})(\rho,\kappa) - \tfrac{v^2}{4}],$$
say for $\gamma = -1/2,$ again as in [4], [5]. Also by (25)
$$\begin{aligned}
g &= A(\rho)d\tau^2 - 4a_0^2\beta^2\kappa^4(sn^2 cn^2)(\rho,\kappa)[a_0^2\kappa^4(\tfrac{sn^2 cn^2}{dn^2})(\rho,\kappa) - \tfrac{v^2}{4}]^{-1} d\rho^2 \\
&= 4a_0^2\beta^2 dn^2(\rho,\kappa)[(a_0^2\kappa^4(\tfrac{sn^2 cn^2}{dn^2})(\rho,\kappa) - \tfrac{v^2}{4})d\tau^2 \\
&\quad -\kappa^4(\tfrac{sn^2 cn^2}{dn^2})(\rho,\kappa)[a_0^2\kappa^4(\tfrac{sn^2 cn^2}{dn^2})(\rho,\kappa) - \tfrac{v^2}{4}]^{-1} d\rho^2],
\end{aligned} \qquad (26)$$
which is exactly the diagonal metric that we focused on in the paper [1]. See definition (6) there, where the notation $a, b$ corresponds to $a_0, 2\beta$ here, respectively, with $v$ there the same as $v$ here – a soliton velocity parameter. We see, as indicated in the introduction, that indeed the consideration of the plasma metric here with $\alpha_1$ in (12) allowed to be *non-zero* can lead to a vast generalization of some of the work in [1].

We turn now to the lingering question of conditions that will imply that $A(\rho) \neq 0,$ and thus validate formula (25). In the special case just considered, for $A(\rho)$ in (26), the single condition
$$\frac{v^2}{4a_0^2\kappa^4} > 1$$
suffices, as shown in [1] – an argument there being based on the inequality
$$\frac{sn^2(x,\kappa)cn^2(x,\kappa)}{dn^2(x,\kappa)} \leq 1, \qquad (27)$$
which we shall use again here. For simplicity of notation we shall write $sn, cn, dn,$ suppressing the variables $\rho, \kappa.$

By equation (23),
$$\begin{aligned}
\frac{-2\gamma A(\rho)}{dn^2} &= 4a_0^2\beta^2[a_0^2\kappa^4(\tfrac{sn\, cn}{dn})^2 - \tfrac{v^2}{4}] - \tfrac{v^2\alpha_1}{4dn^2} + \tfrac{vC}{2\beta dn^2} \\
&\quad - \frac{[16\alpha_1\kappa^4 a_0^4\beta^2(\tfrac{sn\, cn}{dn})^2 + \tfrac{C^2}{\beta^2 dn^2}]}{4[\alpha_1 + 4a_0^2\beta^2 dn^2]}.
\end{aligned} \qquad (28)$$

Multiplication by $4a_0^2\beta^2$ and division by $D \stackrel{def}{=} 4[\alpha_1 + 4a_0^2\beta^2 dn^2] > 0$ lead to 6 terms here. The first and fifth term (in order) combined simplify to
$$4a_0^4\beta^2\kappa^4 \frac{sn^2 cn^2}{dn^2} \frac{4a_0^2\beta^2 dn^2}{\frac{D}{4}},$$
which means that we can write (28) as
$$\begin{aligned}
\frac{-2\gamma A(\rho)}{dn^2} &= 4a_0^2\beta^2\kappa^4(\tfrac{sn^2 cn^2}{dn^2} \cdot \tfrac{4a_0^4\beta^2 dn^2}{\frac{D}{4}}) - a_0^2\beta^2 v^2 - \tfrac{v^2\alpha_1}{4dn^2} \\
&\quad + \tfrac{vC}{2\beta dn^2} - \tfrac{C^2}{D\beta^2 dn^2}.
\end{aligned} \qquad (29)$$
In particular if $A(\rho) = 0$ for some $\rho,$ then
$$\begin{aligned}
v^2(a_0^2\beta^2 + \tfrac{\alpha_1}{4dn^2}) - \tfrac{vC}{2\beta dn^2} + \tfrac{C^2}{D\beta^2 dn^2} &= 4a_0^4\beta^2\kappa^4(\tfrac{sn^2 cn^2}{dn^2} \cdot \tfrac{4a_0^2\beta^2 dn^2}{\frac{D}{4}}) \\
&\leq (4a_0^4\beta^2\kappa^4) \cdot (16a_0^4\beta^2 \tfrac{dn^2}{D}) \\
&\leq 4a_0^4\beta^2\kappa^4
\end{aligned} \qquad (30)$$
by (27), and the fact that $\alpha_1 \geq 0 \Rightarrow D \geq 16a_0^2\beta^2 dn^2.$ That is, (30) says that

$$v^2 a_0^2 \beta^2 \leq -(\frac{v^2 \alpha_1}{4} - \frac{vC}{2\beta}) \frac{1}{dn^2} + 4a_0^4 \beta^2 \kappa^4 - \frac{C^2}{D\beta^2 dn^2}$$
$$\leq -(\frac{v^2 \alpha_1}{4} - \frac{vC}{2\beta}) \frac{1}{dn^2} + 4a_0^4 \beta^2 \kappa^4 \qquad (31)$$

if $A(\rho)$ vanishes at some point $\rho$. Now assume that
$$\frac{v^2 \alpha_1}{4} - \frac{vC}{2\beta} \geq 0.$$
Then $v^2 a_0^2 \beta^2 \leq 4a_0^4 \beta^2 \kappa^4$ by (31) : $v^2 \leq 4a_0^2 \kappa^4$. In other words, if both conditions
$$v^2 > 4a_0^2 \kappa^4 \quad and \quad \alpha_1 v \geq \frac{2C}{\beta} \qquad (32)$$
hold (the first one being the single condition prior to (27) for the special case $\alpha_1 = 0$), then $A(\rho)$ *cannot* vanish at any point $\rho$. Of course if $\alpha_1 = 0$ then (again) $C = 0$ by definition and the second condition in (32) is the triviality $0 \geq 0$. If $\alpha_1 > 0$, then since $\alpha_3 > \alpha_2 \geq \alpha_1, C > 0$ by (12) and
$$v \geq \frac{2C}{\alpha_1 \beta} \Rightarrow v^2 \geq \frac{4C^2}{\alpha_1^2 \beta^2} > 0.$$
That is, conditions for the non-vanishing of $A(\rho)$ are (for $\alpha_1 \neq 0$)
$$v^2 > 4a_0^2 \kappa^4 \quad and \quad v^2 \geq \frac{4C^2}{\alpha_1^2 \beta^2}, \qquad (33)$$
with the single condition $v^2 > 4a_0^2 \kappa^4$ if $\alpha_1 = 0$.

## 4 Mapping the cold plasma metric to a black hole metric

We are in a position now to proceed towards a formulation of the main results. A precise mapping $(\tau, \rho) \to (\tau, r)$ of the plasma metric $g_{plasma}$ in (25) with coordinates $(\tau, \rho)$ to a J-T black hole metric $g_{bh}$, given in (46) below, with coordinates $(\tau, r)$, is presented in equation (48). For a suitable dilaton field $\Phi$, the pair $(g_{bh}, \Phi)$ solves J-T field equations that involve a negative cosmological constant $\Lambda$ that we express (interestingly enough) in terms of $\gamma$ and $\delta$ in the RNLS equation (2), $\delta$ there (as pointed out in the introduction) being the same $\delta$ in the cold plasma system (1). The Hawking black hole temperature and entropy are also expresses in terms of $\gamma$ and $\delta$, and in terms of the parameters $\alpha_1, a_0, v, \kappa$ that describe the plasma density $\rho_1(x,t)$ in (14), and hence also the velocity $u_1(x,t)$ in (12). The mapping is used, moreover, to construct an explicit dilaton field $\Phi_{plasma}$, and we show that the pair $(g_{plasma}, \Phi_{plasma})$ is also a solution of the J-T field equations – a solution in terms of Jacobi elliptic functions, which thus is another extension of results in [1].

Two trivial corrections of errors in [1] are in order:

1. On page 4 there, in the second equation for $ds^2$ in (28) the $dt^2$ should read $d\tau^2$.
2. On page 10 the factor $e^{S/\beta}$ for $s$ in (76) should read $e^{-S/\beta}$.

With (25) now established on solid ground, via the assumptions (33), we first set up a critical change of variables by which $g_{plasma}$ assumes a considerably more manageable form — a form which is a step away from a J-T black hole metric. This key change of variables is given by
$$x \stackrel{def}{=} \frac{a_0^2 \beta^2 \kappa^2}{\gamma} sn^2(\rho, \kappa). \qquad (34)$$

From (16), (34)
$$dx = \frac{2a_0^2\beta^2\kappa^2}{\gamma}(sn\ cn\ dn)(\rho,\kappa)d\rho,$$
$$dx^2 = \frac{4a_0^4\beta^4\kappa^4}{\gamma^2}(sn^2cn^2dn^2)(\rho,\kappa)d\rho^2, \tag{35}$$
$$sn^2(\rho,\kappa) = \frac{\gamma x}{a_0^2\beta^2\kappa^2}, \quad cn^2(\rho,\kappa) = 1 - \frac{\gamma x}{a_0^2\beta^2\kappa^2}, \quad dn^2(\rho,\kappa) = 1 - \frac{\gamma x}{a_0^2\beta^2},$$
which by (25) gives for $A(x) = A(\rho)$ with $x, \rho$ related by (34)
$$g = A(x)d\tau^2 - \frac{dx^2}{A(x)}. \tag{36}$$
Plug the last three elliptic functions expressions in (35) into (23). A modest amount of work renders the result
$$A(x) = \frac{2\gamma x^2}{\beta^2} - (2a_0^2\kappa^2 + \frac{v^2}{2})x + \frac{a_0^2\beta^2 v^2}{2\gamma} + \frac{v^2\alpha_1}{8\gamma} - \frac{vC}{4\gamma\beta}$$
$$+(\frac{-2\alpha_1\gamma x^2}{\beta^2} + 2\alpha_1\kappa^2 a_0^2 x + \frac{C^2}{8\gamma\beta^2})(\frac{1}{-4\gamma x + \alpha_1 + 4a_0^2\beta^2}), \tag{37}$$
which might be seen as limited progress, but the point is the latter quotient term, which we denote by $Q(x)$, very fortunately by long division has a zero remainder. To see this, note that
$$(-4\gamma x + \alpha_3)(x + \frac{\alpha_2}{4\gamma}) = -4\gamma x^2 + (\alpha_3 - \alpha_2)x + \frac{\alpha_2\alpha_3}{4\gamma},$$
where,
$$\alpha_3 - \alpha_2 = \kappa^2(\alpha_3 - \alpha_1)(\text{by equation (12)}) = 4\kappa^2 a_0^2\beta^2\ (\text{by}(13))$$
$$\Rightarrow (\frac{\alpha_1}{2\beta^2})(-4\gamma x + \alpha_3)(x + \frac{\alpha_2}{4\gamma}) = \frac{-2\alpha_1\gamma x^2}{\beta^2} + 2\alpha_1\kappa^2 a_0^2 x + \frac{C^2}{8\gamma\beta^2},$$
which is precisely the numerator $n(x)$ of $Q(x)$ since $C^2 \stackrel{def}{=} \alpha_1\alpha_2\alpha_3$ by (12). However (again by (13)) $\alpha_3 = \alpha_1 + 4a_0^2\beta^2$ and therefore we see also that
$$n(x) = (\frac{\alpha_1}{2\beta^2})(x + \frac{\alpha_2}{4\gamma})d(x)$$
where $d(x)$ is the denominator of $Q(x) \Rightarrow Q(x) = (\frac{\alpha_1}{2\beta^2})(x + \frac{\alpha_2}{4\gamma}) = \frac{\alpha_1 x}{2\beta^2} + \frac{\alpha_1\alpha_2}{8\gamma\beta^2}$, where
$$\alpha_1\alpha_2 = \frac{C^2}{\alpha_3} = \frac{C^2}{(\alpha_1 + 4a_0^2\beta^2)} = \alpha_1[4a_0^2\beta^2(1 - \kappa^2) + \alpha_1]$$
(by (15)) leads to
$$Q(x) = \frac{\alpha_1 x}{2\beta^2} + \frac{\alpha_1}{8\gamma\beta^2}[4a_0^2\beta^2(1 - \kappa^2) + \alpha_1]. \tag{38}$$
This expression substituted in (37) gives the useful result that $A(x)$ is simply a quadratic polynomial in $x$:
$$A(x) = \frac{2\gamma x^2}{\beta^2} - (2a_0^2\kappa^2 + \frac{v^2}{2} - \frac{\alpha_1}{2\beta^2})x + \frac{\alpha_1}{8\gamma\beta^2}[4a_0^2\beta^2(1 - \kappa^2) + \alpha_1]$$
$$+ \frac{a_0^2\beta^2 v^2}{2\gamma} + \frac{v^2\alpha_1}{8\gamma} - \frac{vC}{4\gamma\beta} \tag{39}$$
with $C$ given by (15), as usual.

Equations (36), (39) show that $g_{plasma}$ has the form
$$g = -[A_1 x^2 + B_1 x + C_1]d\tau^2 + [A_1 x^2 + B_1 x + C_1]^{-1}dx^2$$
for $A_1 \stackrel{def}{=} \frac{-2\gamma}{\beta^2} > 0, \quad B_1 \stackrel{def}{=} 2a_0^2\kappa^2 + \frac{v^2}{2} - \frac{\alpha_1}{2\beta^2},$
$$C_1 \stackrel{def}{=} \frac{-\alpha_1}{8\gamma\beta^2}[4a_0^2\beta^2(1 - \kappa^2) + \alpha_1] - \frac{a_0^2\beta^2 v^2}{2\gamma} - \frac{v^2\alpha_1}{8\gamma} \tag{40}$$

$$+\frac{v}{4\gamma\beta}(\alpha_1[4a_0^2\beta^2(1-\kappa^2)+\alpha_1][4a_0^2\beta^2+\alpha_1])^{1/2}.$$

Therefore since $A_1 \neq 0$, it follows directly that the change of variables $r \stackrel{def}{=} x + \frac{B_1}{2A_1}$ transforms $g$ to the J-T Lorentzian form

$$g_{bh} = -[A_1 r^2 + C_1 - \frac{B_1^2}{4A_1}]d\tau^2 + \frac{dr^2}{A_1 r^2 + C_1 - \frac{B_1^2}{4A_1}}. \qquad (41)$$

By (40), one can calculate that

$$\begin{aligned} C_1 - \frac{B_1^2}{4A_1} &= \frac{v}{4\gamma\beta}(\alpha_1[4a_0^2\beta^2(1-\kappa^2)+\alpha_1][4a_0^2\beta^2+\alpha_1])^{1/2} - \frac{a_0^2\beta^2 v^2}{2\gamma} \\ &+ \frac{a_0^4\kappa^4\beta^2}{2\gamma} + \frac{a_0^2\kappa^2 v^2\beta^2}{4\gamma} + \frac{v^4\beta^2}{32\gamma} - \frac{3v^2\alpha_1}{16\gamma} - \frac{3\alpha_1^2}{32\gamma\beta^2} \\ &+ \frac{\alpha_1 a_0^2\kappa^2}{4\gamma} - \frac{\alpha_1 a_0^2}{2\gamma}. \end{aligned} \qquad (42)$$

Before we declare $g_{bh}$ to be an authentic J-T black hole metric, we would first like to have that $C_1 - \frac{B_1^2}{4A_1} < 0$ since, for example, we would look for an event horizon by setting $A_1 r^2 + C_1 - \frac{B_1^2}{4A_1} = 0 : r = \frac{\sqrt{-(C_1 - \frac{B_1^2}{4A_1})}}{\sqrt{A_1}}$, where already $A_1 > 0$ by definition (40), as $\gamma < 0$. We saw that $A(\rho)$ was non-vanishing provided that $v^2$ was sufficiently large; see the conditions in (33). Similarly we check that also $C_1 - \frac{B_1^2}{4A_1} < 0$ provided $v^2$ is sufficiently large; see the conditions in (43), (45) below.

Note that $k \leq 1 \Rightarrow 2\kappa^2 + \kappa^4 \leq 2 + 1 < 4$, in particular. That is
$$2(2 - \kappa^2) > \kappa^4 \Rightarrow 8a_0^2(2-\kappa^2) > 4a_0^2\kappa^4.$$
Thus if $v^2 > 8a_0^2(2-\kappa^2)$, then $v^2 > 4a_0^2\kappa^4$, which is the first condition in (33). Suppose in fact that

$$v^2 > 8a_0^2(2-\kappa^2) + \frac{6\alpha_1}{\beta^2}. \qquad (43)$$

Then since $\gamma < 0$,
$$(\frac{v^2\beta^2}{32\gamma})v^2 < (\frac{v^2\beta^2}{32\gamma})(8a_0^2(2-\kappa^2)) + (\frac{v^2\beta^2}{32\gamma})\frac{6\alpha_1}{\beta^2} = \frac{v^2 a_0^2 \beta^2}{2\gamma} - \frac{v^2 a_0^2\kappa^2\beta^2}{4\gamma} + \frac{3v^2\alpha_1}{16\gamma};$$

that is

$$\frac{v^4\beta^2}{32\gamma} - \frac{v^2 a_0^2\beta^2}{2\gamma} + \frac{v^2 a_0^2\kappa^2\beta^2}{4\gamma} - \frac{3v^2\alpha_1}{16\gamma} < 0. \qquad (44)$$

Therefore to insure that $\frac{C_1 - B_1^2}{4A_1} < 0$ in (42), it is enough to require the condition

$$\frac{vC}{4\gamma\beta} + \frac{a_0^4\kappa^4\beta^2}{2\gamma} - \frac{3\alpha_1^2}{32\gamma\beta^2} + \frac{\alpha_1 a_0^2\kappa^2}{4\gamma} - \frac{\alpha_1 a_0^2}{2\gamma} < 0$$

on the remaining terms there, as one continues to keep (15) in mind:

$$\frac{vC}{4\beta} + \frac{a_0^4\kappa^4\beta^2}{2} - \frac{3\alpha_1^2}{32\beta^2} + \frac{\alpha_1 a_0^2\kappa^2}{4} - \frac{\alpha_1 a_0^2}{2} > 0$$

happens in particular if

$$\frac{vC}{4\beta} - \frac{3\alpha_1^2}{32\beta^2} - \frac{\alpha_1 a_0^2}{2} \geq 0,$$

which is automatic if $\alpha_1 = 0$:
$$v \geq \frac{\alpha_1}{C}[\frac{3\alpha_1}{8\beta} + 2a_0^2\beta], \quad \alpha_1 \neq 0 \tag{45}$$

With conditions (43), (45) in place to guarantee that $C_1 - \frac{B_1^2}{4A_1} < 0$, we express $g_{bh}$ in (41) as a J-T black hole metric in the form
$$g_{bh} = -[m^2 r^2 - M]d\tau^2 + \frac{dr^2}{m^2 r^2 - M} \tag{46}$$
for $m^2 = A_1 = \frac{-2\gamma}{\beta^2} > 0$, $M = -[C_1 - \frac{B_1^2}{4A_1}] > 0$, M being a mass parameter given by (42).
We have noted that (43) implies, in particular, the first condition $v^2 > 4a_0^2\kappa^4$ in (33).
The composite change of variables $(\tau,\rho) \to (\tau,x) \to (\tau,r)$ therefore transforms the cold plasma metric $g_{plasma}$ in (25) to the black hole metric $g_{bh}$ in (46). This composition is immediately computed: $(\tau,\rho) \to (\tau,r)$ where (by (34), (40), (16))

$$r \stackrel{def}{=} x + \frac{B_1}{2A_1} \stackrel{def}{=} \frac{a_0^2\beta^2\kappa^2}{\gamma}sn^2(\rho,\kappa) + \frac{2a_0^2\kappa^2 + \frac{v^2}{2} - \frac{\alpha_1}{2\beta^2}}{\frac{-4\gamma}{\beta^2}}$$
$$= \frac{a_0^2\beta^2}{\gamma}[1 - dn^2(\rho,\kappa)] - \frac{a_0^2\beta^2\kappa^2}{2\gamma} - \frac{v^2\beta^2}{8\gamma} + \frac{\alpha_1}{8\gamma}. \tag{47}$$

That is,
$$r = \psi(\rho) \stackrel{def}{=} \frac{a_0^2\beta^2 dn^2(\rho,\kappa)}{-\gamma} + \frac{a_0^2\beta^2}{2\gamma}[2-\kappa^2] + \frac{v^2\beta^2 - \alpha_1}{-8\gamma}. \tag{48}$$

Going back to (25) again, which we write as
$$g_{plasma} = A(\rho)d\tau^2 + \frac{B(\rho)}{A(\rho)}d\rho^2,$$
$$B(\rho) \stackrel{def}{=} \frac{-4a_0^4\beta^4\kappa^4}{\gamma^2}(sn^2 cn^2 dn^2)(\rho,\kappa), \tag{49}$$

we note that $B(\rho)$ and $\psi(\rho)$ in (48) are in fact related:
$$[\psi'(\rho)]^2 = -B(\rho), \tag{50}$$
a fact that will be useful later.

$g_{bh}$ solves the J-T field equations
$$\begin{aligned} R(g) + 2\Lambda &= 0 \\ \nabla_i \nabla_j \Phi + \Lambda \Phi g_{ij} &= 0, \end{aligned} \tag{51}$$
where $\Lambda$ is a *negative* cosmological constant, $\Phi$ is a *dilaton* field, and where the Hessian is given by
$$\nabla_i \nabla_j = \frac{\partial^2}{\partial x_i \partial x_j} - \sum_{k=1}^{2}\Gamma_{ij}^k \frac{\partial}{\partial x_k}, \tag{52}$$
for the (second) Christoffel symbols $\Gamma_{ij}^k$ of $g$. In the present case $(x_1,x_2) = (\tau,r)$ and
$$\Lambda = -m^2 \stackrel{def}{=} \frac{2\gamma}{\beta^2} \stackrel{def}{=} \frac{2\gamma}{(\delta-1)}, \quad \Phi(\tau,r) \stackrel{def}{=} mr. \tag{53}$$
The equations of motion (51) are derived from the J-T action integral
$$I_{J-T}(g,\Phi) = \frac{1}{2G}\int_{M^2} d^2x\sqrt{|det\ g|}\ \Phi\ (R(g) + 2\Lambda) \tag{54}$$
for a two-dimensional spacetime $M^2$ and a gravitational coupling constant $G$.

For $\hbar = \frac{h}{2\pi}$ the normalized Planck constant, one has the general formulas for the black hole Hawking temperature $T_H$ and Bekenstein-Hawking entropy $S_{BH}$ [14]:

$$T_H = \frac{\hbar G}{\pi}\sqrt{M} = \frac{\hbar G}{\pi}\sqrt{-(C_1 - \frac{B_1^2}{4A_1})},$$
$$S_{BH} = \frac{2\pi}{\hbar G}\sqrt{M} = \frac{2\pi}{\hbar G}\sqrt{-(C_1 - \frac{B_1^2}{4A_1})}.$$
(55)

In particular, $T_H$ and $S_{BH}$ are also expressed in terms of solution data for the plasma system (1).

Given $\psi, \Phi$ in (48),(53), define $\Phi_{plasma}$ by
$$\Phi_{plasma}(\tau, \rho) = \Phi(\tau, \psi(\rho)). \tag{56}$$

The Christoffel symbols for $g_{plasma}$ and $g_{bh}$ (in the form expressed in (49) and (46)) can be computed using Maple, for example. The non-vanishing ones are given, respectively, by

$$\Gamma^1_{12} = \frac{1}{2}\frac{A'(\rho)}{A(\rho)}, \quad \Gamma^2_{11} = \frac{-1}{2}\frac{A(\rho)A'(\rho)}{B(\rho)}, \quad \Gamma^2_{22} = \frac{1}{2}[\frac{B'(\rho)}{B(\rho)} - \frac{A'(\rho)}{A(\rho)}], \tag{57}$$
$$\Gamma^1_{12} = \frac{-m^2 r}{-m^2 r^2 + M}, \quad \Gamma^2_{11} = -(-m^2 r^2 + M)m^2 r, \quad \Gamma^2_{22} = \frac{m^2 r}{-m^2 r^2 + M}.$$

By our notation/definition, $A(\rho) = -m^2\psi(\rho)^2 + M \Rightarrow A'(\rho) = -2m^2\psi(\rho)\psi'(\rho)$, which permits one to relate the Christoffel symbols of $g_{plasma}$ and $g_{bh}$ in (57):

$$\Gamma^1_{12}(\tau, \rho) = \Gamma^1_{12}(\tau, \psi(\rho))\psi'(\rho)$$
$$\Gamma^2_{11}(\tau, \rho) = \frac{-\Gamma^2_{11}(\tau, \psi(\rho))\psi'(\rho)}{B(\rho)} \tag{58}$$
$$\Gamma^2_{22}(\tau, \rho) = \frac{1}{2}\frac{B'(\rho)}{B(\rho)} + \Gamma^2_{22}(\tau, \psi(\rho))\psi'(\rho),$$

the left hand side of these equations being the Christoffel symbols of $g_{plasma}$. Using the equations in (58), we can, in the end, relate the Hessian acting on $\Phi_{plasma}$ and $\Phi$ in (56), (53); see (52):

$$(\nabla_1\nabla_1\Phi_{plasma})(\tau, \rho) = (\nabla_1\nabla_1\Phi)(\tau, \psi(\rho))$$
$$(\nabla_1\nabla_2\Phi_{plasma})(\tau, \rho) = (\nabla_1\nabla_2\Phi)(\tau, r) = 0 \tag{59}$$
$$(\nabla_2\nabla_2\Phi_{plasma})(\tau, \rho) = -B(\rho)(\nabla_2\nabla_2\Phi_{plasma})(\tau, \psi(\rho)),$$

where for the last formula here we use (50) and its implication $2\psi'(\rho)\psi''(\rho) = -B'(\rho)$.

Since the pair $(g_{bh}, \Phi)$ solves the J-T field equations (51) (as we have noted, for the cosmological constant $\Lambda = -m^2 \stackrel{def}{=} \frac{2\gamma}{\beta^2}$), one concludes from (59) that the pair $(g_{plasma}, \Phi_{plasma})$ also solves the J-T field equations, where by (48), (56) $\Phi_{plasma}$ is given explicitly by

$$\Phi_{plasma}(\tau, \rho) = m[\frac{a_0^2\beta^2 dn^2(\rho, \kappa)}{-\gamma} + \frac{a_0^2\beta^2(2-\kappa^2)}{2\gamma} + \frac{(v^2\beta^2 - \alpha_1)}{-8\gamma}]. \tag{60}$$

For convenience, we iterate that the plasma metric for the dilaton field $\Phi_{plasma}$ in (60) is given by equations (23), (25):

$$g_{plasma} = A(\rho)d\tau^2 - [\frac{4\kappa^4 a_0^4 \beta^4}{A(\rho)\gamma^2}(sn^2 cn^2 dn^2)(\rho, \kappa)]d\rho^2 \tag{61}$$

for

$$A(\rho) = 4a_0^2 \beta^2 \left[-\frac{a_0^4 \kappa^4}{2\gamma}(sn^2\ cn^2)(\rho,\kappa) + \frac{v^2}{8\gamma} dn^2(\rho,\kappa)\right]$$

$$+ \frac{16\alpha_1\beta^2\ \kappa^4 a_0^4 (sn^2\ cn^2)(\rho,\kappa) + \frac{C^2}{\beta^2}}{8\gamma[\alpha_1 + 4a_0^2\beta^2\ dn^2(\rho,\kappa)]} + \frac{v^2\alpha_1}{8\gamma} - \frac{vC}{4\gamma\beta},$$

(62)

$$C = \alpha_1^{1/2}\ [4a_0^2\beta^2(1-\kappa^2) + \alpha_1]^{1/2}\ [4a_0^2\beta^2 + \alpha_1]^{1/2}\ .$$

In the special, but important, case when the elliptic modulus $\kappa$ is chosen to be 1, $\Phi_{plasma}$ in (60) reduces to the equation

$$\Phi_{plasma}(\tau,\rho) = m\left[\frac{a_0^2\beta^2 sech^2\rho}{-\gamma} + \frac{a_0^2\beta^2}{2\gamma} + \frac{(v^2\beta^2-\alpha_1)}{-8\gamma}\right] \quad (63)$$

and $g_{plasma}$ in (61) simplifies to

$$g_{plasma} = A(\rho)d\tau^2 - \left[\frac{4a_0^4\beta^4}{A(\rho)\gamma^2}(tanh^2 sech^4)(\rho)\right]d\rho^2 \quad (64)$$

for

$$A(\rho) = 4a_0^2\ \beta^2 \left[-\frac{a_0^4}{2\gamma}(tanh^2 sech^2)(\rho) + \frac{v^2}{8\gamma} sech^2(\rho)\right]$$

$$+ \frac{16\alpha_1\beta^2\ a_0^4(tanh^2 sech^2)(\rho) + \frac{C^2}{\beta^2}}{8\gamma[\alpha_1 + 4a_0^2\beta^2\ sech^2(\rho)]} + \frac{v^2\alpha_1}{8\gamma} - \frac{vC}{4\gamma\beta},$$

(65)

$$C = \alpha_1(4a_0^2\beta^2 + \alpha_1)^{1/2},$$

by (16). Lastly as the only restriction on $\alpha_1$ is $\alpha_1 \geq 0$, one can certainly specialize the choice $\alpha_1 = 0$ in (63), in which case $C = 0$ and $A(\rho)$ in (64), (65) simplifies to

$$A(\rho) = 4a_0^2\ \beta^2 \left[-\frac{a_0^4}{2\gamma}(tanh^2 sech^2)(\rho) + \frac{v^2}{8\gamma} sech^2(\rho)\right]. \quad (66)$$

Also for the choices $\alpha_2 = \alpha_1 = 1$ (relevant for the plasma physics) we have noted in formula (19) the simpler expression for the Gurevich–Krylov traveling wave solution of the plasma density $\rho_1(x,t)$ (and hence of its velocity $u_1(x,t)$ by (12)) in terms of the hyperbolic secant.

## 5  Closing Remarks

The transformation formula (48) is the key connection between the cold plasma metric $g_{plasma}$

in (25) attached to the solutions (12), (14) of the magnetoacoustic system (1) and the J-T black hole metric $g_{bh}$ in (46) - a connection nicely facilitated by the general correspondence set up in Section 2 between solutions of (1) and of the resonance Schrödinger equation (2). The dilaton field $\Phi_{plasma}$ constructed in (60) provides for a solution $(g_{plasma}, \Phi_{plasma})$ of the J-T field equations (51) (in terms of Jacobi elliptic functions), where the cosmological constant $\Lambda$ is negative and is expressed in (53) in terms of the parameters $\gamma, \delta$ in the Schrödinger equation (2), $\gamma$ being assumed negative (which was exploited in Section 2 in the aforementioned correspondence of solutions of (1) and (2), and $\delta > 1$ being the same $\delta$ in (1)).

The conditions established in (43), (45) insure positivity of the black hole mass M in (46). In particular, the expressions in (55) for Hawking temperature and entropy are meaningful. However, one could relax these conditions and obtain, more generally, by way of (48) a connection between cold plasma and 2d black holes with a naked singularity (for $M < 0$) and black hole vacua, with $M = 0$. It should be possible, extending methods presented here, to construct additional dilaton fields $\Phi$ such that the pairs $(g_{plasma}, \Phi)$ also solve the J-T field equations (51) – this and other matters the author could consider in future work.

The construction of the plasma metric (61), which eventually was transformed to the black hole metric (46), was based on solutions of the RDS (5) generated by solutions of the RNLS equation (2), or equivalently (as shown in Section 2) by solutions of the MAS (1) – given the general prescription (6). Apart from the traveling wave, one – soliton solutions of (2), or of (5), that arise here for $\kappa = 1$, one could consider multi-soliton solutions as well. In Section 6 of [3], Lee and Pashaev apply the Hirota bilinear representation to construct, for example, an explicit two – dissipaton solution of the RDS (5). This means that in principle one could construct the corresponding constant scalar curvature metric (again by way of (6)), attempt to diagonalize it, and map the diagonalized form to a black hole metric. Obviously that task would be more difficult than that which was dealt with here, but it would be an extremely interesting, fascinating task to pursue as it would likely provide insight, for example, on the collision of black holes in the plasma, or possibly multi-black hole solutions in J-T gravity.

## Conflicts of Interest

The author declares that there are no conflicts of interest regarding the publication of this paper.

## Acknowledgments

This paper draws inspiration from the work of the various authors in [4], [7]. We thank Mr. Ojaswin Karthikeyan for his excellent preparation of the manuscript.## References

[1]   J. D'Ambroise and F. Williams, *"Elliptic function solutions in Jackiw-Teitelboim dilaton gravity",* Advances in Mathematical Physics, vol. 2017, Article ID 2154784, Hindawi, 2017.

[2]   A. Gurevich and A. Krylov, *"A shock wave in dispersive hydrodynamics",* Soviet

Phys. Dokl. vol 32, pp. 73-74, 1988.

[3]   J.-H. Lee and O. Pashaev, *"Resonance NLS solitons as black holes in Madelung fluid",* Modern Physics Letters A, vol. 17, no.24, pp. 1601-1619, 2002.

[4]   J.-H. Lee, O. Pashaev, C. Rogers, and W. Schief, *"The resonant nonlinear Schrödinger equation in cold plasma physics. Applications of Bäcklund-Darboux transformations and superposition principles",* Journal of Plasma Physics, vol. 73, no. 2, pp. 257-272, 2007.

[5]   J.-H. Lee and O. Pashaev, *"Solitons of the resonant nonlinear Schrödinger equation with nontrivial boundary conditions and Hirota bilinear method",* Theoretical and Mathematical Physics, vol. 152, no. 1, pp. 991-1003, 2007.

[6]   J.-H. Lee, O. Pashaev, C. Rogers, *"Soliton resonances in a generalized nonlinear Schrödinger equation",* Journal of Physics A, Math. and Theor., vol. 41, no. 45, Article ID 452001, 9 pages. 2008.

[7]   L. Martina, O. Pashaev, and G. Soliani, *"Integrable dissipative structures in the gauge theory of gravity",* Classical and Quantum Gravity, vol. 14, no. 12, pp. 3179-3186, 1997.

[8]   L. Martina, O. Pashaev, and G. Soliani, *"Bright solitons as black holes",* Physical Review D, vol. 58, no.8, Article ID 084025, 13 pages, 1998.

[9]   R. Jackiw, *"A two-dimensional model of gravity",* in Quantum Theory of Gravity, S. Christensen, Ed., pp. 403-420, Adam Hilger Ltd., 1984.

[10]  C. Teitelboim, *"The Hamiltonian structure of two-dimensional spacetime and its relation with the conformal anomaly",* in Quantum Theory of Gravity, S. Christensen, Ed., pp. 327-344, Adam Hilger Ltd., 1984.

[11]  T. Klösch and T. Strobl, *"Classical and quantum gravity in 1+1 dimensions. Part 1: A unifying approach",* Classical and Quantum Gravity, vol. 13, pp. 965-984, 1996; Erratum-ibid vol.14, no.3, pp. 875, 1997.

[12]  F. Williams, *"Some selected thoughts old and new on soliton–black hole connections in 2d dilaton gravity",* in The Sine-Gordon Model and its Applications: From Pendula and Josephson Junctions to Gravity and High-Energy Physics, J. Cuevas-Maraver, P. Kevrekidis, and F. Williams, Eds., pp. 177-205, Springer Publisher, 2014.

[13]  K. Chandrasekharan, *"Elliptic Functions",* Grundlehren der mathematischen Wissenschaften, vol. 281, Springer–Verlag Publisher, 1985.

[14]  J. Lemos, *"Thermodynamics of the two-dimensional black hole in the Jackiw-Teitelboim theory",* Physical Review D, vol. 54, no. 10, pp. 6206-6212, 1996.